\documentclass[letterpaper, 10 pt, conference]{ieeeconf} 
\IEEEoverridecommandlockouts   % This command is only needed if you want to use the \thanks command

\makeatletter
\def\endthebibliography{%
  \def\@noitemerr{\@latex@warning{Empty `thebibliography' environment}}%
  \endlist
}
\makeatother

\usepackage{cite}
\usepackage{amsmath,amssymb,amsfonts}
\usepackage{algorithmic}
\usepackage{textcomp}
\usepackage{xcolor}
\def\BibTeX{{\rm B\kern-.05em{\sc i\kern-.025em b}\kern-.08em
    T\kern-.1667em\lower.7ex\hbox{E}\kern-.125emX}}
\usepackage{amssymb}
\let\labelindent\relax % Clash with enumitem
\usepackage{enumitem}
\usepackage{amsthm}% Formats (bold, no indent) theorems, comments and stuff
\newtheoremstyle{dotless}{}{}{\itshape}{}{\bfseries}{}{ }{}

\usepackage{amsmath}
\usepackage{graphicx}
\graphicspath{{./Figures}}
\DeclareGraphicsExtensions{.pdf,.jpeg,.png}

\usepackage{tikz} % Caption legends
\usetikzlibrary{decorations.pathreplacing, patterns, calc, shapes,shapes.misc}
\tikzset{cross/.style={cross out, draw=black, minimum size=2*(#1-\pgflinewidth), inner sep=0pt, outer sep=0pt}, tcross/.default={1pt}}

\renewcommand{\baselinestretch}{0.97} 
\makeatletter
\g@addto@macro\normalsize{%
	\setlength\abovedisplayskip{1.2ex plus 1pt minus 1pt}
	\setlength\belowdisplayskip{\abovedisplayskip}
	\setlength\abovedisplayshortskip{0pt plus 1 pt}
	\setlength\belowdisplayshortskip{\abovedisplayskip}
}
\makeatother
\definecolor{modelcolor}{rgb}{0,0.72202,0}
\definecolor{FRFcolor}{rgb}{0,0,0}

\definecolor{withRC}{rgb}{0.5176,0.8235,0}
\definecolor{withoutRC}{rgb}{0,0,1}

\newcommand{\markerFRF}{\raisebox{.5ex}{\tikz{\draw[FRFcolor, line width = 0.4mm] (0,0) -- +(1em, 0);}}}

\newcommand{\markerwithRCtimestamps}{\raisebox{.2ex}{\tikz{\draw[withRC] (0,0) node[cross=2.5pt,withRC] {};}}}
\newcommand{\markerwithoutRCtimestamps}{\raisebox{.3ex}{\tikz{\draw[draw=withoutRC] (0,0) circle (0.4ex) ;}}}

\newcommand{\linewithRC}{\raisebox{.3ex}{\tikz{\draw[withRC, line width = 0.4mm] (0,0) -- +(1em, 0);}}}
\newcommand{\linewithoutRC}{\raisebox{.3ex}{\tikz{\draw[withoutRC, line width = 0.4mm] (0,0) -- +(1em, 0);}}}
\newcommand{\linedashed}{\raisebox{.3ex}{\tikz{\draw[black, line width = 0.4mm, dashed] (0,0) -- +(1em, 0);}}}
\newcommand{\linesolid}{\raisebox{.3ex}{\tikz{\draw[black, line width = 0.4mm, solid] (0,0) -- +(1em, 0);}}}
\newcommand{\linedotted}{\raisebox{.3ex}{\tikz{\draw[black, line width = 0.4mm, dotted] (0,0) -- +(1em, 0);}}}

\newcommand{\markerlearningcurvewithRC}{\raisebox{.3ex}{\tikz{\draw[draw=withRC,fill=withRC] (0,0) circle (0.4ex) ;}}}
\newcommand{\markerlearningcurvewithoutRC}{\raisebox{.3ex}{\tikz{\draw[draw=withoutRC,fill=withoutRC] (0,0) circle (0.4ex) ;}}}

\definecolor{Infnorm}{rgb}{1,0,1}
\definecolor{MAnorm}{rgb}{1,0.5020,0}
\definecolor{RMSnorm}{rgb}{0,0,0}

\newcommand{\markerRMSnorm}{\raisebox{.5ex}{\tikz{\draw[RMSnorm, line width = 0.4mm] (0,0) -- +(1em, 0);}}}

\theoremstyle{dotless}
\newtheorem{dummy}{}
\newtheorem{assumption}[dummy]{Assumption}
\newtheorem{theorem}[dummy]{Theorem}

\newtheorem{lemma}[dummy]{Lemma}

\newtheorem{example}[dummy]{Example}
\newtheorem{definition}[dummy]{Definition}

\newtheorem{procedure}[dummy]{Procedure}

\newcommand{\norm}[2]{\lVert#1\rVert_{#2}}
\newcommand{\abs}[1]{\left\lvert #1 \right\rvert}
\newcommand{\ejw}{e^{j\omega}}
\newcommand{\ejwmin}{e^{-j\omega}}
\newcommand{\rhinf}{\mathcal{RH}_{\infty}}

\begin{document}

\bstctlcite{IEEEexample:BSTcontrol} % Fix ---- for author names in bibliography

\title{\LARGE \bf Intermittent Sampling in Repetitive Control:\\ Exploiting Time-Varying Measurements
}

\author{Johan Kon, Nard Strijbosch, Sjirk Koekebakker, and Tom Oomen% <-this % stops a space
\thanks{Johan Kon, Nard Strijbosch and Tom Oomen are with the Control Systems Technology
Group, Dept. of Mechanical Engineering, Eindhoven University
of Technology, Eindhoven, The Netherlands, e-mail: j.j.kon@tue.nl,
n.w.a.strijbosch@tue.nl, t.a.e.oomen@tue.nl. Tom Oomen is also with the Delft Center for Systems and Control, Delft University of Technology, Delft, The Netherlands. Sjirk Koekebakker is with Canon Production Printing, Venlo, The Netherlands, e-mail: sjirk.koekebakker@cpp.canon. This
work is part of the research programme VIDI with project number 15698,
which is (partly) financed by the Netherlands Organisation for Scientific
Research.}% <-this % stops a space
}

\maketitle

\begin{abstract}
The performance increase up to the sensor resolution in repetitive control (RC) invalidates the standard assumption in RC that data is available at equidistant time instances, e.g., in systems with package loss or when exploiting timestamped data from optical encoders. The aim of this paper is to develop an intermittent sampling RC framework for non-equidistant measurements. Sufficient stability conditions are derived that can be verified using non-parametric frequency response function data. This results in a frequency domain design procedure to explicitly address uncertainty. The RC framework is validated on an industrial printbelt setup for which exact non-equidistant measurement data is available.
\end{abstract}

\section{Introduction}
Learning control methods, including repetitive control (RC) \cite{LONGMAN2010447} and iterative learning control (ILC) \cite{1636313}, can achieve major performance improvements up to the sensor resolution for periodic tasks \cite{Longman1998, 8891752}. RC and ILC are based on the internal model principle (IMP) \cite{FRANCIS1976457}, in which the key idea is to iteratively adjust the input based on errors from the previous executions of the same task by including a model of the disturbance in the controller. The main difference between ILC and RC is that the final state of the current task is carried over to the initial state of the next task in RC. For ILC, the state is reset after each task to the same initial condition.
% such that the initial state is task-invariant.

Successful RC methods typically rely on 1) a stability test that can be verified using non-parametric models in the form of identified frequency response function (FRF) data and 2) an intuitive frequency domain design framework based on loop-shaping to explicitly address uncertainty \cite{doi:10.1080/002071700405905, 248004}. A common approach is based on a buffer in positive feedback interconnection with itself \cite{10.1115/1.3153060, 1274}, complemented by a learning and robustness filter, with possible extensions to uncertain period times \cite{STEINBUCH20022103, STEINBUCH20072086} and multiple periods \cite{8962172,912291}. Further refinements include basis functions in the form of parallel periodic signal generators \cite{Shi_2014}. These frameworks rest on the assumption that exact data is available at all samples, allowing for linear time-invariant (LTI) stability and design frameworks such as Nyquist and Bode diagrams.

The assumption in RC that exact measurement data is available equidistantly in time, i.e., at all sample instances, is not always valid. For example, encoders sampled at an equidistant rate result in quantization errors. However, they are exact at the timestamp of the measurement \cite{b9aabc96e9bb48989fda9003b2110edd}. Package loss in networked control systems similarly results in intermittently sampled measurements. In low-performance feedback control, intermittent sampling effects and resulting quantization can be neglected as they are insignificant compared to the magnitude of the converged error. Due to the increased performance up to the sensor resolution in RC, these effects are no longer negligible and, if left unaddressed, deteriorate performance.

% The increase in performance up to the level of sensor resolution by applying RC challenges the fundamental assumption that exact measurement data is available at all sample instances, i.e., that it is equidistant in time. For example, measurement data from sensors with finite resolution such as encoders is only exact at line transitions \cite{8815329} and thus non-equidistant in time, i.e., only sampled intermittently. Package loss in networked control systems similarly results in intermittently sampled measurements. Due to the increased performance of RC, these effects are no longer negligible and if left unaddressed, result in quantization, which is not modelled by the internal model, thus decreasing performance. 

Non-equidistant measurement data is addressed in ILC through combining the IMP with a worst-case analysis for arbitrary time-varying measurement points. Alternatively, the availability of data at each time instant is modelled as a known probability distribution in \cite{AHN200812442,doi:10.1080/00207179.2014.986762}. This probability distribution is unknown in many applications, thus, a worst-case analysis is preferred. However, none of these methods can straightforwardly be extended to RC due to the state-reset in ILC that is not present in RC, necessitating an infinite-time instead of finite-time analysis.

Although RC has been substantially developed for a large range of practical problems, at present an RC approach that can deal with intermittent measurement, as occurring in, e.g., encoders or networked control systems, is not yet available. The aim of this paper is to design an RC framework that can asymptotically reject periodic disturbances when measurement data is available at non-equidistant samples.

The main contribution of this paper is an intermittent sampling RC framework that employs intermittent data for asymptotically rejecting periodic disturbances that can be successfully implemented in practice. This is achieved through the following sub-contributions. 
\begin{enumerate}[label=C\arabic*]
	\item Stability tests for any realization of the measurement data, based on (identified) FRF data (Section \ref{sec:stability}).
	\item Design guidelines through loop-shaping techniques to explicitly address uncertainty (Section \ref{sec:design}).
	\item Experimental validation of the developed framework on an industrial printbelt setup (Section \ref{sec:experimental_results}).
\end{enumerate}

%\begin{enumerate}[label=C\arabic*]
%	\item Stability tests for any realization of the measurement data, which can be verified based on identified FRF data (Section \ref{sec:stability}).
%	\item Guidelines for design through frequency domain loop-shaping techniques to explicitly address uncertainty (Section \ref{sec:design}).
%	\item Experimental validation of the developed framework on an industrial printbelt setup (Section \ref{sec:experimental_results}).
%\end{enumerate}

\subsection*{Notation and Definitions}
All systems are discrete-time, single-input single-output (SISO) and linear with sample rate $T_s$. The set of real rational, SISO, proper, asymptotically input-output stable, discrete-time transfer functions is denoted by $\rhinf$. The sets $\mathbb{N}$ and $\mathbb{Z}$ indicate the set of all positive integers, and the set of all integers, respectively. The $\ell_2$-norm of an infinite sequence of scalars $y = \left(\ldots, y(-2), y(1), y(0), y(1), y(2), \ldots\right)$, $y(k) \in \mathbb{R}$, $k \in \mathbb{Z}$, is defined as $\norm{y}{\ell_2} \doteq \left( \sum_{k=-\infty}^\infty \abs{y(k)}^2 \right)^{\frac{1}{2}}$. The space $\ell_2$ consists of all infinite sequences with finite $\ell_2$-norm.

\section{Problem Formulation}
In this section, the problem of intermittent sampling RC is introduced. First, traditional RC is presented. Second, the concept of timestamped data is introduced and combined with the traditional RC setup. Lastly, the problem is posed.
 
\subsection{Repetitive Control}
\begin{figure}[t]
	\centering
	\includegraphics[]{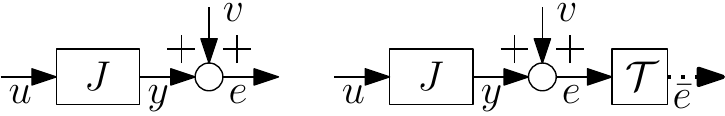}
	\caption{Traditional repetitive control setup (left) and intermittent sampling repetitive control setup (right) with periodic disturbance $v$ and timestamping operator $\mathcal{T}$ resulting in the intermittently sampled error $\bar{e}$. Equidistant signals are indicated by full lines (\protect \linesolid), intermittent signals are dotted (\protect \linedotted).}
	\label{fig:RC_setup}
	\vspace{-1.0em}
\end{figure}

RC is a feedback control method to asymptotically reject periodic disturbances acting on a dynamic process. Consider the repetitive control setup in Fig. \ref{fig:RC_setup} (left). The goal of RC is to reject the influence of the periodic disturbance $v(k) \in \mathbb{R}$ on the error $e(k) \in \mathbb{R}$ given by
\begin{align}
	e(k) &= y(k) + v(k),
\end{align} 
where $y(k) \in \mathbb{R}$ is the output of dynamic process $J$, i.e.,
\begin{align}
	y(k) &= J u(k).
\end{align}
Here $u(k) \in \mathbb{R}$ is the input to the process, and $k \in \mathbb{Z}$ the discrete timestep. $J$ can represent either a closed-loop or open-loop system. The following is assumed.
\begin{assumption}
	The dynamic process $J(z) \in \rhinf$ with $J(z)$ strictly proper, i.e., $\lim_{z \rightarrow \infty} J(z) = 0$.
	\label{ass:plant_stable_and_strictlyproper}
\end{assumption}

\noindent The disturbance $v$ is periodic with period $N$ according to
\begin{equation}
	v(k+N) = v(k) \ \forall k.
\end{equation}

\begin{figure}[t]
	\centering
	\includegraphics[]{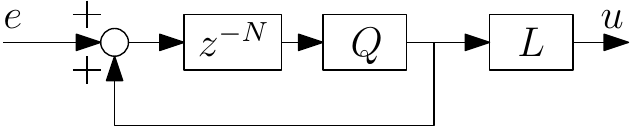}
	\caption{Typical repetitive controller with buffer $z^{-N}$ in positive feedback interconnection as periodic signal generator, learning filter $L$, and cut-off filter $Q$.}
	\label{fig:standard_RC}
	\vspace{-1.5em}
\end{figure}

The internal model principle (IMP) \cite{FRANCIS1976457} states that an internal model of the disturbance has to be included in a feedback controller to asymptotically reject the influence of the disturbance on the dynamic process, i.e.,
\begin{equation}
	\lim_{k \rightarrow \infty} e(k) = 0.
\end{equation} 
The structure of a typical RC is depicted in Fig. \ref{fig:standard_RC}. In this RC, the internal model is formed by a buffer with length $N$ in positive feedback interconnection with itself, complemented by learning filter $L$ and robustness filter $Q$ according to
\begin{equation}
	\frac{u(z)}{e(z)} = R(z) = \frac{LQ z^{-N}}{1 - Q z^{-N}},
	\label{eq:standard_RC}
\end{equation}
where $z^{-N}$ represents the buffer, and $z^{-n_L} L, z^{-n_Q} Q \in \rhinf$, with finite preview $n_L, n_Q \in \mathbb{Z}_{\geq 0}$. 
%\textcolor{red}{Alternatively, the internal model can be parametrized by $M$ parallel periodic signal generators with a distinct frequency $f_m$ \cite{Shi_2014}, or by cascaded repetitive controllers with distinct buffer lengths \cite{8962172}, resulting in different realizations of $R(z)$}.

A large class of repetitive controllers can be designed in the frequency domain using non-parametric models, more specifically, using identified FRF data. The repetitive controller of Fig. \ref{fig:standard_RC} is stable for all buffer lengths $N$ if
\begin{equation}
	\sup\nolimits_{\omega \in [0,\pi]}  \abs{\left(1-J(\ejw)L(\ejw)\right)Q(\ejw)} < 1,
\label{eq:nominal_stability_standard_RC}
\end{equation}
see \cite{LONGMAN2010447}. The above criterion can directly be verified using non-parametric FRF data for $J(\ejw)$. Consequently, $L$ is designed as an (approximate) inverse of $J$ for stability. $Q$ is designed as a low-pass filter to prevent learning instability for high frequencies due to mismatch of the plant inverse. Multiple methods exist for designing the approximate inverse $L$, such as ZPETC \cite{10.1115/1.3143822} or preview control \cite{VANZUNDERT2018282}. $Q$ is usually designed as a zero-phase finite impulse response filter \cite{LONGMAN2010447}.

%Repetitive control has been successfully applied in practice to reduce periodic errors to the level of sensor resolution in numerous applications \cite{doi:10.1080/002071700405905}. Most of these applications have been in an LTI setting. Dit moet naar intro
In this traditional setting, it is assumed that error information is available at each timestep $k$. There exists a well established RC framework to reject the disturbance $v$ based on this equidistant error data.

\subsection{The Timestamping Operator}
In many control applications, a measurement of $e(k)$ is not available at each timestep, but the time instance of the measurement, i.e., its timestamp, is available, resulting in non-equidistant but exact measurement data. This can be modelled by a timestamping operator that encodes the availability of information at each timestep. The timestamping operator is a memoryless, linear time-varying (LTV) system.
\begin{definition}
The timestamping operator $\mathcal{T}$ outputs its input at the timestamps, and 0 otherwise, according to
\begin{equation}
	\mathcal{T}: e(k) \rightarrow \bar{e}(k),\ \bar{e}(k) = \begin{cases} e(k) & \mathrm{if}\ k \in \Psi \\ 0 & \mathrm{otherwise}, \end{cases}
	\label{eq:timestamper}
\end{equation}
in which $\Psi$ is a sequence of timestamps, unknown in advance but measurable during operation, according to
\begin{equation}
	\Psi = \{\bar{k}_1, \bar{k}_2, \ldots \},\ \bar{k}_i \in \mathbb{N}, \ \bar{k}_i < \bar{k}_{i+1},
\end{equation}
where $\bar{k}_i$ is the timestamp of measurement $i \in \mathbb{Z}_{> 0}$.
\end{definition}

\noindent In the RC setting with a timestamping operator, only $\bar{e}(k)$ is available to reject periodic disturbances $v(k)$. To reject $v(k)$ in this setting, the intermittent sampling repetitive control setup in Fig. \ref{fig:RC_setup} (right) is considered. Two assumptions are made regarding the timestamping operator.
\begin{assumption}
The measurement of the error is exact at the timestamps, i.e., $e(k) = y(k) + v(k) \ \forall k \in \Psi$.
\label{ass:exact_at_timestamps}
\end{assumption}

\begin{assumption}
The sequence of timestamps $\Psi$ is independent of the timestamping operator input $e$. 
\label{ass:random_timestamper}
\end{assumption}
\noindent Assumption \ref{ass:exact_at_timestamps} defines the ideal timestamping operator. Assumption \ref{ass:random_timestamper} allows for stability guarantees through a worst-case analysis, in which the sequence of timestamps $\Psi$ consists of arbitrary non-equidistant timestamps. 
%Consequently, the stability guarantees are independent of the realization of $\Psi$, and thus valid for any system described by the timestamping operator. 

The timestamping operator can be used to model a variety of systems, e.g.,
\begin{enumerate}
	\item optical encoders that output the exact position at line transitions \cite{8815329}, and
	% and quantized data otherwise. 
	%If only the exact data at the line transitions is used by the repretitive controller and quantized data is disregarded as in\cite{b9aabc96e9bb48989fda9003b2110edd}, the timestamper can be used to model optical encoders.
	\item networked control systems subject to package loss.
\end{enumerate} 
\noindent %In both situations, if no new measurement is communicated at timestep $k$, the repetitive controller can reuse the measurement of timestep $k-1$, or no data can be used at all. The latter situation is modelled by the timestamper. 
Low-resolution optical encoders fit directly in this framework, since the measurement can be considered exact at the timestamps. High-resolution encoders require a high sampling rate of the encoder output. This can be captured in the framework by exploiting frequency domain lifting techniques \cite{9385661} to obtain a single-rate setting.

The problem considered is the design of repetitive controllers for process $J$ using the intermittent error $\bar{e}$ from timestamping operator $\mathcal{T}$.

\subsection{Problem Definition}
The aim of this paper is to develop an intermittent sampling repetitive control framework to asymptotically reject the influence of $v(k)$ on dynamic process $J$ using intermittent samples $\bar{e}(k)$ from timestamping operator $\mathcal{T}$. This includes
\begin{enumerate}
	\item developing frequency domain stability tests for arbitrary timestamp realizations $\Psi$ based on non-parametric FRF data analogous to \eqref{eq:nominal_stability_standard_RC} (Section \ref{sec:stability}),
	\item converting the stability test into a systematic frequency domain design framework that is similar to existing design frameworks (Section \ref{sec:design}), and
	\item validating the framework experimentally (Section \ref{sec:experimental_results}).
\end{enumerate}

\section{Stability of the Intermittent Sampling Repetitive Control Setup} 
\label{sec:stability}
In this section, stability criteria for the intermittent sampling RC setup of Fig. \ref{fig:RC_setup}, that can be validated using identified FRF data, are derived. First, the intermittent sampling RC setup is rewritten to the standard negative feedback interconnection of Fig. \ref{fig:intermittent_sampling_RC_transformed2} (left). Secondly, two stability conditions are derived, constituting contribution C1. 

%The main result of this section is a passivity based stability test that can be validated using measured FRF data, and this result is compared to a small-gain stability analysis.

\subsection{The Timestamping Operator Complement}
\begin{figure}[t]
	\centering
	\includegraphics[width=\linewidth]{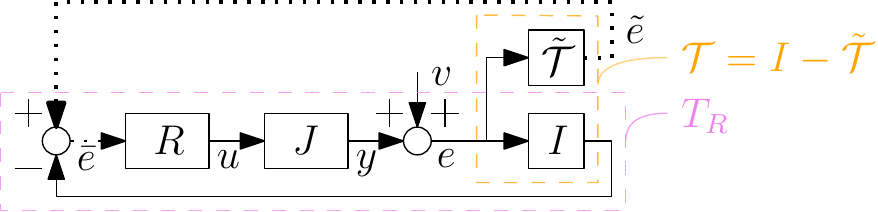}
	\caption{Intermittently sampled repetitive control setup as double feedback interconnection of $RJ$ with LTI path (lower) with unity gain and LTV path (upper) with timestamping operator complement $\tilde{\mathcal{T}}$.}
	\label{fig:intermittent_sampling_RC_transformed}
	\vspace{-1.5em}
\end{figure}

\begin{figure}[t]
	\centering
	\includegraphics[]{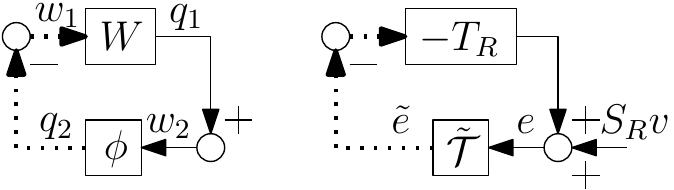}
	\caption{Standard negative feedback interconnection with discrete time LTI system $W$ and memoryless function $\phi$ (left). The intermittent sampling RC setup with LTI negative complementary sensitivity $-T_R$ and LTV memoryless timestamping operator complement $\tilde{\mathcal{T}}$ (right) fits in the standard negative feedback interconnection with $W = -T_R$ and $\phi = \tilde{\mathcal{T}}$.}
	%Intermittently sampled RC setup as negative  feedback interconnection of LTI negative nominal complementary sensitivity $-T_R$ and LTV timestamping operator complement $\tilde{\mathcal{T}}$ (left). The intermittently sampled RC setup fits in the standard negative feedback interconnection with system $W \in \rhinf$ and sector bound memoryless function $\phi$ (right) with $-T_R = W$ and $\tilde{\mathcal{T}} = \phi$.
	\label{fig:intermittent_sampling_RC_transformed2}
	\vspace{-1.5em}
\end{figure}

To derive stability results, the intermittent sampling RC setup is rewritten to the standard negative feedback interconnection of Fig. \ref{fig:intermittent_sampling_RC_transformed2} (left), consisting of a discrete time LTI system $W$ and a memoryless function $\phi$.

The intermittent error $\bar{e}(k)$ can be represented as the combination of the error of the traditional RC setup $e(k)$, and a mechanism that models the unavailability of data outside the timestamps, as opposed to the availability of data at the timestamps, see Fig. \ref{fig:intermittent_sampling_RC_transformed}.
%More specifically, the timestamping operator $\mathcal{T}$ is split into identity and its complement that models the unavailability of data,  
%The contribution of $I$ corresponds to the error $e(k)$ of the traditional RC setting, whereas the  complement $\tilde{\mathcal{T}}$ guarantees $\bar{e}(k) = 0$ in case no new data is available, modelling the intermittent sampling as an additional feedback path on top of the traditional setting.
The timestamping operator complement $\tilde{\mathcal{T}}$ is a memoryless, LTV operator that models the unavailability of data as defined below.
\begin{definition}
The timestamping operator complement $\tilde{\mathcal{T}}$ outputs 0 at the timestamps, and its input otherwise, i.e.,
 \begin{equation}
	\tilde{\mathcal{T}}: e(k) \rightarrow \tilde{e}(k), \ \tilde{e}(k) =
	\begin{cases} 0 & \mathrm{if}\ k \in \Psi \\ e(k) & \mathrm{otherwise}.
\end{cases} 
	\label{eq:timestampercomplement}
\end{equation}
such that $\mathcal{T}(k, e(k)) + \tilde{\mathcal{T}}(k, e(k)) = e(k)$.
%$\mathcal{T}(k, e(k)) = e(k) - \tilde{\mathcal{T}}(k,e(k))$
\end{definition}

\noindent Consequently, the intermittent error is given by
\begin{equation}
	\bar{e}(k) = e(k) - \tilde{\mathcal{T}}(k, e(k)),
\end{equation}
such that the intermittent sampling RC setup can be interpreted as a double feedback interconnection, see Fig. \ref{fig:intermittent_sampling_RC_transformed}, in which the error $\bar{e}(k)$ consists of an LTI and equidistantly sampled contribution $e(k)$ and an additional LTV and non-equidistantly sampled contribution $\tilde{\mathcal{T}}(k,e(k))$.

Subsequently, the memoryless function $\tilde{\mathcal{T}}$ is isolated. This is possible due to linearity of the repetitive controller $R$ and dynamic process $J$. The input $e(k)$ of the timestamping operator complement $\tilde{\mathcal{T}}$ is written as a function of its output $\tilde{e}(k)$ and the exogenous signal $v(k)$, see Fig. \ref{fig:intermittent_sampling_RC_transformed}, as
\begin{subequations}
\begin{align}
	e(k) &= T_R \tilde{e}(k) + S_R v(k), \\
	T_R &= \left(1+JR\right)^{-1}JR, \ S_R = \left(1+JR\right)^{-1},
\end{align}
\end{subequations}
resulting in the negative feedback interconnection of $-T_R$ and $\tilde{\mathcal{T}}$ depicted in Fig. \ref{fig:intermittent_sampling_RC_transformed2} (right). $T_R$ can be recognized as the complementary sensitivity of the RC loop in the traditional setting. As the timestamping operator complement is the only time-varying operator, the resulting systems $T_R$, $S_R$ are LTI. 

The resulting interconnection fits in the standard negative feedback interconnection of Fig. \ref{fig:intermittent_sampling_RC_transformed2} (left) with discrete time LTI system $W = -T_R$ and memoryless function $\phi = \tilde{\mathcal{T}}$, and enables the application of stability analysis based on dissipativity and small-gain theory, as investigated next.

\subsection{Passivity-based Stability Analysis}
In this section, stability of the intermittent sampling RC setup using passivity arguments is established, for a given RC design $R$ and based on identified FRF data of $J$. This constitutes contribution C1. 

Consider again the standard negative feedback interconnection of Fig. \ref{fig:intermittent_sampling_RC_transformed2} (left). The following result applies.

\begin{lemma}
The real rational transfer function $W(z) \in \rhinf$ is discrete positive real (DPR) if
	\begin{enumerate}
		\item $W(z)$ is analytic in $\abs{z} \geq 1$, and
		\item $W(\ejw) + W(\ejwmin) = 2\Re(W(\ejw)) \geq 0 \ \forall \omega \in [0,\pi]$.
	\end{enumerate}
	\label{lem:DPR}
\end{lemma}
\noindent Lemma \ref{lem:DPR} is a special case of Lemma 2 of \cite{5248618}, with $W$ SISO and $W \in \rhinf$. The following Lemma relates uniform stability of the standard negative feedback interconnection of Fig. \ref{fig:intermittent_sampling_RC_transformed2} to a DPR condition on the transfer function $W$.
\begin{lemma}[]
	Consider the negative feedback interconnection of $W$ and $\phi(k,w_2(k))$ in Fig. \ref{fig:intermittent_sampling_RC_transformed2}, where $W$ is strictly proper, and $\phi(k,w_2(k))$ is a memoryless function in the sector with boundaries $q_2 = 0$ and $q_2 = K^{-1} w_2$, $K \in \mathbb{R}_{> 0}$, such that $\forall k \in \mathbb{Z}$ it holds that
\begin{subequations}
\begin{align}
	\phi(k,0) &= 0, \label{eq:sector_bound1} \\
	\phi\left(k,w_2(k)\right) \left(K\phi\left(k,w_2(k)\right) - w_2(k)\right) &\leq 0. \label{eq:sector_bound2}
\end{align}
\label{eq:sector_bound}
\end{subequations}
\hspace{-2ex} If $K+W(z)$ is discrete positive real as in Lemma \ref{lem:DPR}, the feedback system is globally uniformly stable \cite{5248618}.
	\label{lem:DPR_interconnection}
\end{lemma}
\noindent The proof is based on a discrete time counterpart of the Kalman–Yakubovich–Popov lemma. 
%relating discrete positive real transfer functions to linear matrix inequalities, such that there exists a positive definite storage function that is non-increasing along trajectories of the system.
%All required lemmas are now available to develop a stability test for the intermittently sampled RC setup that can be validated using identified FRF data, which constitutes contribution C2.
Lemma \ref{lem:DPR_interconnection} is applied to the timestamping operator complement $\tilde{\mathcal{T}}$.

\begin{lemma}
	The timestamping operator complement $\tilde{\mathcal{T}}$ satisfies the sector bound conditions \eqref{eq:sector_bound} with $K=1$.
	\label{lem:sector_bound_timestamper}
\end{lemma}

Lemma \ref{lem:DPR}, \ref{lem:DPR_interconnection} and \ref{lem:sector_bound_timestamper} allow for the following stability theorem for the intermittent sampling RC setup.

\begin{figure}[t]
	\centering
	\includegraphics[]{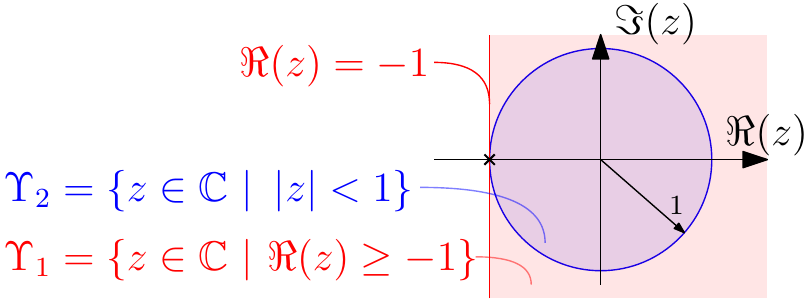}
	\caption{Stability regions of Theorem \ref{th:passivity_based_stability} ($\Upsilon_1$) and Theorem \ref{th:small_gain_stability} ($\Upsilon_2$). If the frequency response of the negative complimentary sensitivity $-T_R(\ejw)$ is confined to either region $\forall \omega \in [0,\pi]$, and $T_R \in \rhinf$, the intermittent sampling repetitive control setup is stable. As $\Upsilon_2 \subset \Upsilon_1$, Theorem \ref{th:small_gain_stability} implies Theorem \ref{th:passivity_based_stability} and introduces extra conservatism.}
	\label{fig:stability_region}
	\vspace{-1.5em}
\end{figure}

\begin{theorem}[Passivity-based stability analysis]
	Consider the intermittent sampling RC setup of Fig. \ref{fig:RC_setup} with $J \in \rhinf$ and $J$ strictly proper, periodic disturbance $v$, and timestamping operator $\mathcal{T}$. The repetitive controller $R$ results in a uniformly stable feedback interconnection if
	\begin{enumerate}[label=S\arabic*)]
		\item $J(\ejw)R(\ejw)$ does not encircle and does not intersect the point $z=-1$ for $\omega \in [0,2\pi)$, and
		\item the negative complementary sensitivity $-T_R(\ejw) = -\left(1 +J(\ejw)R(\ejw)\right)^{-1}J(\ejw)R(\ejw)$ is confined to the region $\Upsilon_1 = \{z \in \mathbb{C} \ \vert \ \Re(z) \geq -1\}$ for $\omega \in [0,\pi]$, see Fig. \ref{fig:stability_region}.
	\end{enumerate}
	\label{th:passivity_based_stability}
\end{theorem}

Theorem \ref{th:passivity_based_stability} is a two-fold stability test based on sequential loop-closing that can be validated using identified FRF data of $J$. It consists of a Nyquist stability requirement S1 for the traditional RC loop based on equidistant error data, and a non-strict passivity requirement S2 as an additional condition imposed by the intermittent sampling characteristics.

\begin{example}
For the typical RC of Fig. \ref{fig:standard_RC} with transfer function \eqref{eq:standard_RC}, the complementary sensitivity reads as 
\begin{equation}
	T_r(z) = \frac{J(z)L(z)Q(z) z^{-N}}{1 - (1-J(z)L(z))Q(z)z^{-N}}.
\end{equation}
Condition S1 of Theorem \ref{th:passivity_based_stability} is implied by the small-gain stability criterion \eqref{eq:nominal_stability_standard_RC} \cite{LONGMAN2010447}, such that $T_R \in \rhinf$. For this RC structure, Condition S2 is given by
\begin{align}
	\Re \left( -\frac{J(\ejw)L(\ejw)Q(\ejw) e^{-j\omega N}}{1 - \left(1-J(\ejw)L(\ejw)\right)Q(\ejw)e^{-j\omega N}} \right)
	\geq -1, \label{eq:IO_stability_standard_RC}
\end{align}
$\forall \omega \in [0, \pi]$, such that the typical RC of Fig. \ref{fig:standard_RC} is stable in the intermittent sampling setting if \eqref{eq:nominal_stability_standard_RC} and \eqref{eq:IO_stability_standard_RC} are satisfied.
\end{example}
 
%\begin{remark}
%For the typical repetitive controller \eqref{eq:standard_RC} of Fig. \ref{fig:standard_RC}, the sufficient small-gain stability condition \eqref{eq:nominal_stability_standard_RC} implies the necessary and sufficient Nyquist condition S1 in Theorem \ref{th:passivity_based_stability}, i.e., it implies $T_R \in \rhinf$.
%\end{remark}

\subsection{Small-gain Stability Analysis}
In this section, small-gain arguments are used to derive a sufficient test for asymptotic stability of the intermittent sampling RC setup, for a given RC design $R$ and based on identified FRF data of $J$. In contrast to the passivity based test, the small-gain test can be validated based as a simple magnitude constraint on the complimentary sensitivity $T_R$, but introduces additional conservatism, see Fig. \ref{fig:stability_region}.
%The result is related to the passivity based analysis of Theorem \ref{th:passivity_based_stability}.

To employ small-gain arguments for stability, the worst-case energy gain of a system, i.e., its induced $\ell_2$-gain, is defined.
\begin{definition}
	The induced $\ell_2$-gain of a linear SISO system $W \in \rhinf$ is given by
	\begin{equation}
		\norm{W}{\ell_2,i} = \sup_{w_1 \in \ell_2, w_1 \neq 0} \frac{\norm{W w_1}{\ell_2}}{\norm{w_1}{\ell_2}}.
	\end{equation}
\end{definition}

\noindent The worst-case energy gain can be used to infer stability of the standard negative feedback interconnection of a linear system $W$ and memoryless function $\phi$. 
\begin{lemma}[Small-gain theorem]
	Consider the negative feedback interconnection of $W$ and $\phi(k,w_2(k))$ in Fig. \ref{fig:intermittent_sampling_RC_transformed2}, where $W$ is strictly proper and $\phi(k,w_2(k))$ is a memoryless function. If the loop gain $W\phi$ satisfies
\begin{equation}
	\norm{W \phi}{\ell_2,i} < 1,
\end{equation}
then the feedback system is globally asymptotically stable.
\label{lem:small_gain_theorem}
\end{lemma}
\noindent See, e.g., \cite{DESOER197536} for a proof. Lemma \ref{lem:small_gain_theorem} allows for the following stability theorem for the intermittent sampling RC setup.

\begin{theorem}[Small-gain stability analysis]
Consider the intermittent sampling RC setup of Fig. \ref{fig:RC_setup} with $J \in \rhinf$ and $J$ strictly proper, with periodic disturbance $v$, and timestamping operator $\mathcal{T}$. The repetitive controller $R$ results in an asymptotically stable feedback interconnection if
	\begin{enumerate}[label=S\arabic*)]
		\item $J(\ejw)R(\ejw)$ does not encircle and does not intersect the point $z=-1$ for $\omega \in [0,2\pi)$, and
		\item the complementary sensitivity $T_R(\ejw) = \left(1 + J(\ejw) \right.$ $\left. R(\ejw)\right)^{-1}J(\ejw)R(\ejw)$ is confined to the region $\Upsilon_2 = \{z \in \mathbb{C} \ \vert \ \abs{z} < 1\}$, i.e., $\abs{T_R(\ejw)} < 1$ for $\omega \in [0,\pi]$ see Fig. \ref{fig:stability_region}.
	\end{enumerate}
	\label{th:small_gain_stability}
\end{theorem}

Similar to Theorem \ref{th:passivity_based_stability}, Theorem \ref{th:small_gain_stability} is a two-fold stability test based on sequential loop-closing that can be validated using identified FRF data of $J$, in which the second stability condition S2 is a small-gain condition on $T_R$ as opposed to a discrete positive real condition on $I-T_R$.

The stability definition of Theorem \ref{th:small_gain_stability} differs from Theorem \ref{th:passivity_based_stability}: for any timestamp realization $\Psi$, asymptotic stability requires convergence, while the passivity argument just requires boundedness. In case no data is available, i.e., $\tilde{\mathcal{T}} = 1$ and $\Psi = \emptyset$, asymptotic stability requires the internal model states, e.g., the buffer elements in the typical RC of Fig. \ref{fig:standard_RC}, to converge. This corresponds to the difference in stability regions, see Fig. \ref{fig:stability_region}, i.e., the difference between the \textit{closed} half space $\Upsilon_1$, which includes the point $z = -1$, and the \textit{open} unit disk, which does not. In other words, Theorem \ref{th:small_gain_stability} implies Theorem \ref{th:passivity_based_stability}.

In conclusion, the small-gain based stability theorem is a more intuitive stability condition that can be validated based on a simple magnitude constraint on $T_R$, but it introduces conservatism with respect to Theorem \ref{th:passivity_based_stability}.

\section{Design Aspects}
\label{sec:design}
In this section, a design framework for intermittent sampling RC is presented that develops intuitive design guidelines to satisfy Theorem \ref{th:passivity_based_stability}, constituting contribution C2.

The design procedure is two-fold. First, the repetitive controller is designed for stability of $T_R$ according to one of the methods in literature \cite{LONGMAN2010447, Shi_2014, doi:10.1080/002071700405905}. Second, the design is validated against the second condition of Theorem \ref{th:passivity_based_stability}. The design is iterated until both conditions are satisfied. When iterating the design, the waterbed effect is often the limiting factor: a sufficiently high learning gain increases the magnitude of $-T_R(\ejw)$ outside $\Upsilon_1$, i.e., left of the line $\Re(z) = -1$, around the cross-over frequency $\omega_0$ for which $\abs{J(e^{j\omega_0})R(e^{j\omega_0})} = 1$.

The following procedure illustrates the design for the typical RC of Fig. \ref{fig:standard_RC}.

\vspace{3pt}
\hrule
\vspace{-2pt}
\begin{procedure}[RC design for intermittent sampling] \hfill
	\vspace{3pt} \hrule
	\begin{enumerate}
		\item Choose the appropriate RC type and determine $T_R$.

		\item Design the RC such that it is stable in the non-intermittent setting, i.e., such that $T_R \in \rhinf$.
		
		For the typical RC of Fig. \ref{fig:standard_RC}, this consists of the following three steps to satisfy \eqref{eq:nominal_stability_standard_RC}.
		\begin{enumerate}
			\item Obtain an approximate parametric model $\hat{J}$.
			\item Design $L$ as an approximation of $\hat{J}^{-1}$ \cite{VANZUNDERT2018282}.
			\item Design $Q$ as a zero-phase finite impulse response low-pass filter such that $Q(\ejw) = 1$ at frequencies where $\abs{I-J(\ejw)L(\ejw)} \ll 1$ for learning performance, and $Q(\ejw) \ll 1$ at frequencies where $\abs{1-J(\ejw)L(\ejw)} \geq 1$ for stability.
		\end{enumerate}
		\item Modify the design for the non-intermittent setting such that S2 of Theorem \ref{th:passivity_based_stability} is satisfied, i.e., such that $-T_R(\ejw) \in \Upsilon_1$, by decreasing the magnitude response of $T_R(\ejw)$ around the cross-over frequency $\omega_0$.
		
		For the typical RC of Fig. \ref{fig:standard_RC}, two design heuristics typically allow for satisfying \eqref{eq:IO_stability_standard_RC}.
		\begin{enumerate}
			\item Scale $L$ as $\alpha L$, $\alpha \in [0,1] \subset \mathbb{R}$, and decrease $\alpha$.
			\item Decrease the magnitude of $Q$ locally around $\Omega = \{\omega \ \vert \ \Re(-T_R(\ejw)) < -1 \}$.
		\end{enumerate} 
	\end{enumerate}
	\label{proc:intermittent_RC_design}
	\vspace{0pt} 	\hrule 	\vspace{-2pt}
\end{procedure}
Both design heuristics in the third step guarantee stability of the RC in the intermittent sampling setting. The first heuristic slows down learning but does not limit the converged learning performance, and is thus preferred over the second, which limits the converged learning performance for frequencies $\omega \in \Omega$, but does not slow down learning.
		
This design procedure can be applied to other RC types, like MBFRC \cite{Shi_2014}, by adhering to the non-intermittent design guidelines for that type in the second step. In the third step, design heuristic 3a can be applied to every RC type.

\section{Experimental Results}
\label{sec:experimental_results}
In this section, the developed RC framework is validated on an industrial printbelt system (contribution C3).

\subsection{Setup}
\begin{figure}[t]
	\includegraphics[width=\linewidth]{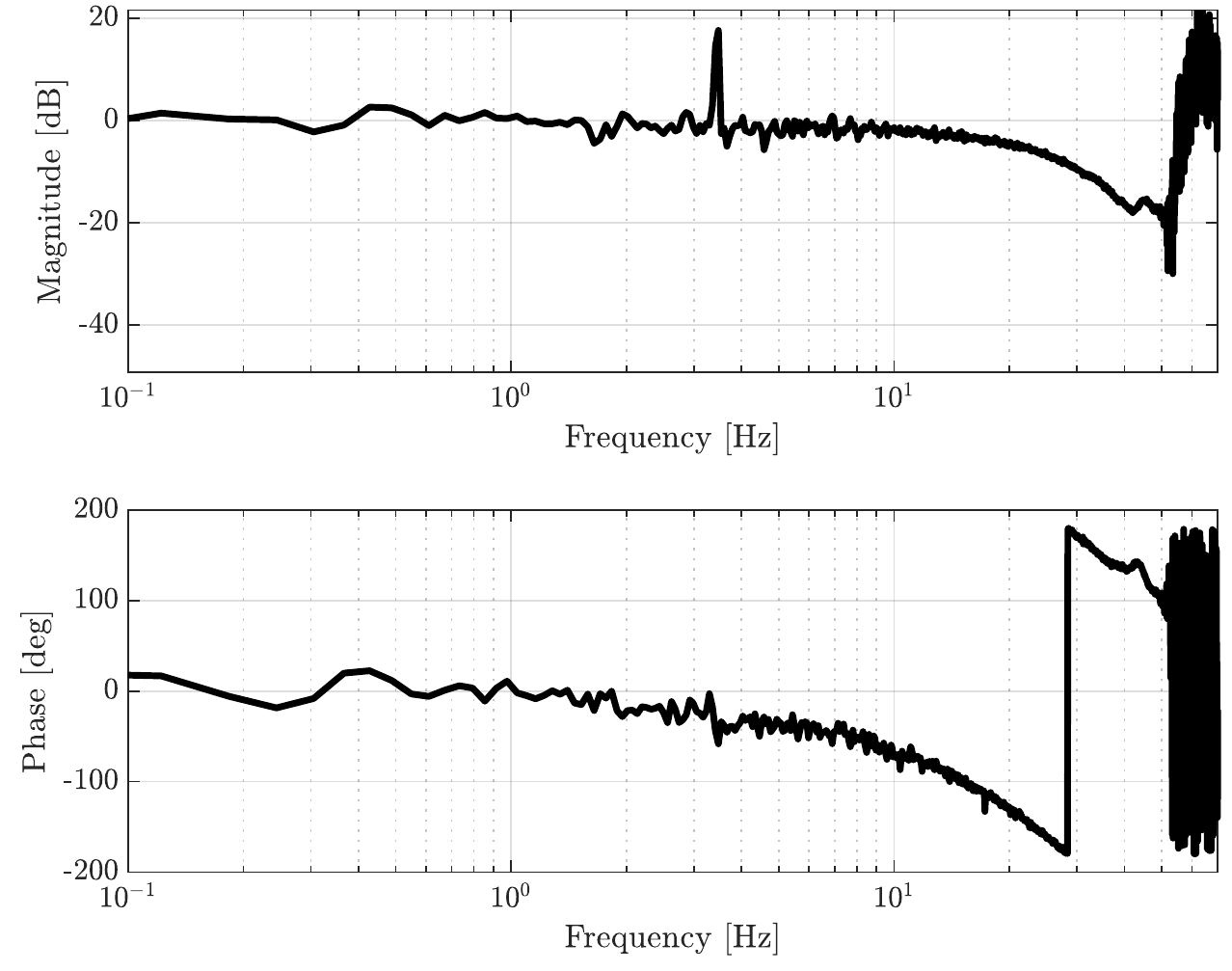}
	\vspace{-2em}
	\caption{Bode diagram of identified FRF (\protect \markerFRF) of transfer function $J$ from repetitive controller output $u$ to output $y$ used for MBFRC design. Note that no information is available above 50 Hz due to the limited power of the excitation signal above this frequency. The peak in magnitude around 3.5 Hz is caused by a drivetrain disturbance at this frequency, resulting in extra output power.
	}
	\label{fig:plant_FRF}
	\vspace{-1.5em}
\end{figure}

The framework is illustrated using a printbelt module similar to the one in the Canon Varioprint i300 \cite{VPi300}. In this printer, paper sheets are transported underneath an inkjet printing station by a printbelt, which is guided along a set of rollers. The belt is actuated by a voltage-driven motor at 500 Hz. The position of the paper is measured by means of a perforated belt edge and optical sensors. This printbelt setup, in continuous operation with constant velocity, fits in the intermittent sampling framework of Fig. \ref{fig:RC_setup} (right), as the distance between the holes in the belt is large, such that measurements of the belt position are intermittent and can be considered exact at the timestamps. The Bode diagram of the transfer from motor input to belt position (pre-stabilized by feedback on another sensor) is shown in Fig. \ref{fig:plant_FRF}. Note that advanced system identification techniques are necessary due to the intermittent sampling characteristics \cite{LARSSON2002709}.
%Note that no data is available above $50$ Hz, as no input power was applied in the identification above this frequency: the approximate sampling rate of the intermittent belt measurements is $110$ Hz.

\subsection{Experimental Results}
The main disturbance is caused by the drivetrain dynamics and eccentricity with a period that is a non-integer amount of samples. Consequently, a matched basis functions repetitive controller (MBFRC) \cite{Shi_2014} is designed to reject this disturbance. The MBFRC is designed such that the conditions in Theorem \ref{th:passivity_based_stability} are satisfied based on the identified FRF data of $J$ without obtaining a parametric model, guaranteeing stability of the intermittently sampled repetitive control setup.

\begin{figure}[!ht]
	\includegraphics[width=\linewidth]{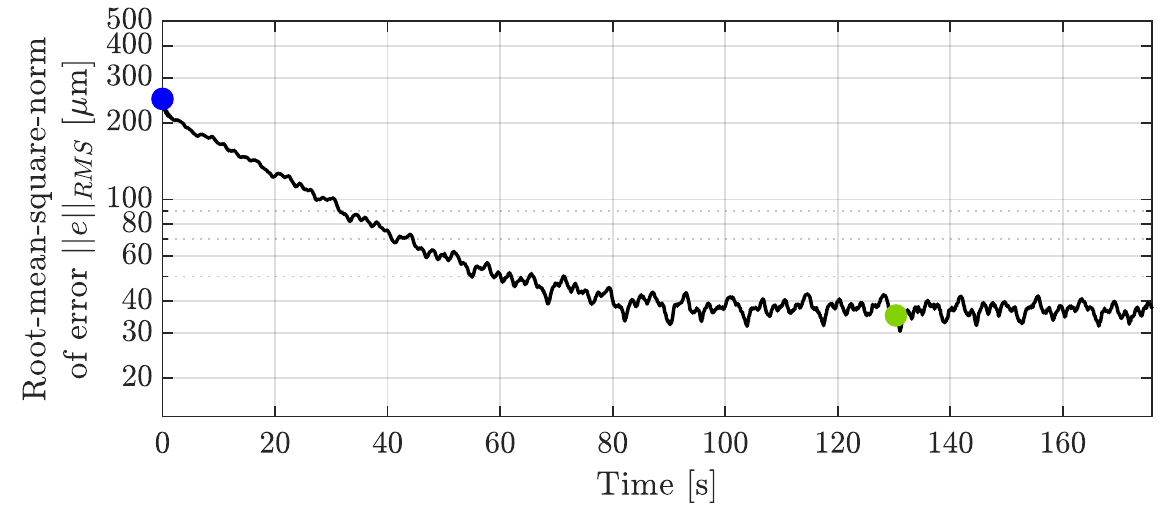}
	\vspace{-2em}
	\caption{Root-mean-square-norm $\norm{e}{RMS} = \sqrt{\frac{1}{N} \sum_{n = k - N }^{n=k} e(n)^2}$ (\protect \markerRMSnorm) of intermittent sampling MBFRC error (linearly interpolated to equidistant time grid) in moving window of length $2.9$ s. The $RMS$-norm of the error is reduced by a factor 7 by the MBFRC design. The time and cumulative amplitude spectrum of the initial (\protect \markerlearningcurvewithoutRC) and converged (\protect \markerlearningcurvewithRC) error are shown in Fig. \ref{fig:error_time} and \ref{fig:error_freq}.}
	%Learning curve of the intermittent sampling MBFRC design as a function of the experiment time. A trial consists of one period of the most dominant disturbance, i.e. the inverse of the drivetrain frequency. The mean absolute error $\norm{e}{MA} = \frac{1}{K} \sum_{k=1}^K \abs{e(k)}$ (\protect \markerMAnorm), root-mean-square error $\norm{e}{RMS} = \sqrt{\frac{1}{K} \sum_{k=1}^K e(k)^2}$ (\protect \markerRMSnorm), and infinity norm of the error $\norm{e}{\infty} = \max_{k \in [1,K]} \abs{e(k)}$ (\protect \markerInfnorm) are reduced by a factor 6, 6.5 and 5 respectively.}
	\label{fig:error_learning_curve}
\end{figure}

Fig. \ref{fig:error_learning_curve} shows the learning curve of the repetitive controller. The MBFRC is able to reduce the $RMS$-norm of the error by a factor 7 based on only the intermittently sampled non-equidistant error, as compared to a situation without RC.

\begin{figure}[!ht]
	\vspace{-1em}
	\includegraphics[width=\linewidth]{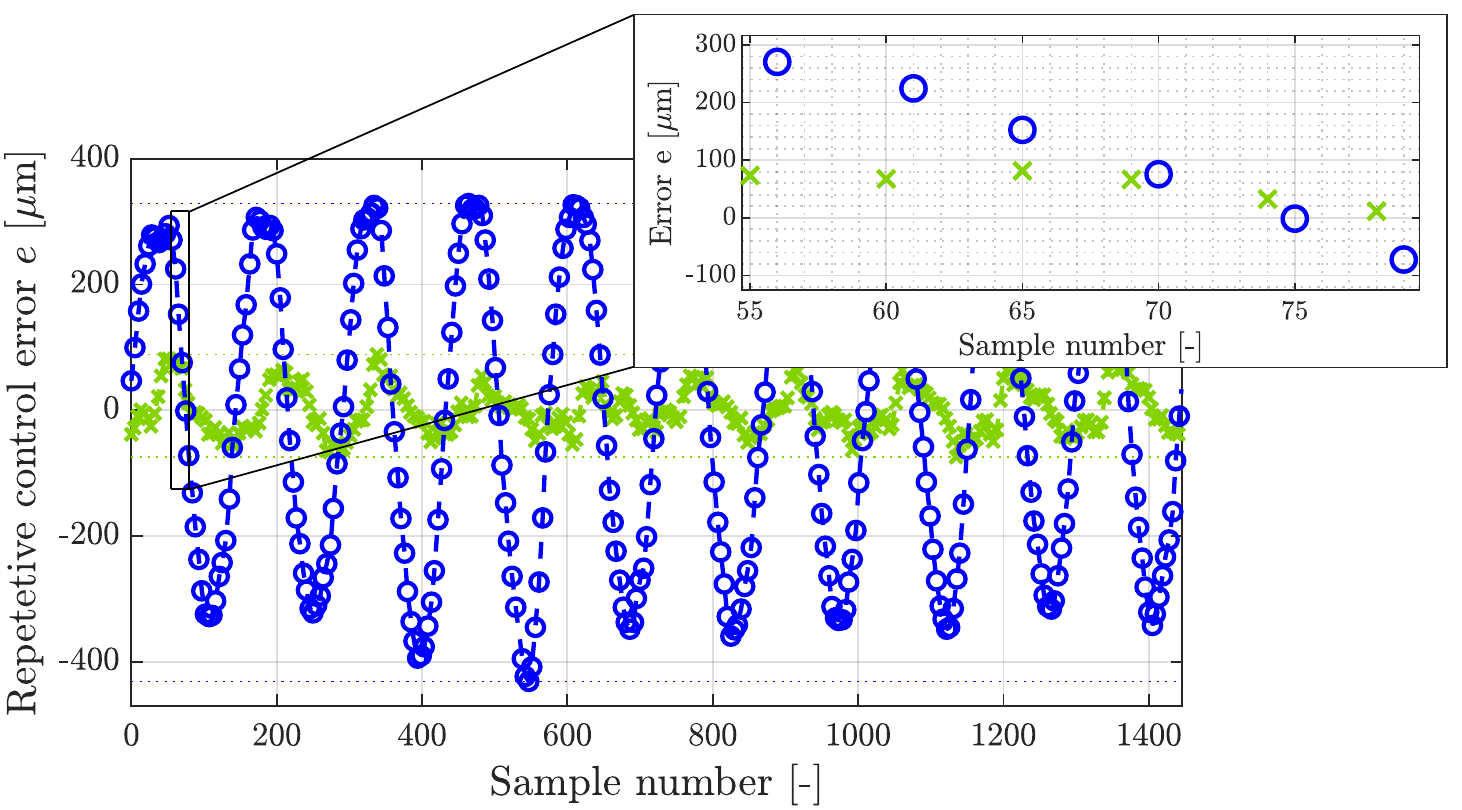}
	\vspace{-2em}
	\caption{Converged error of the intermittent sampling MBFRC design compared to the initial error. The error at the timestamps is given by (\protect \markerwithRCtimestamps) and (\protect \markerwithoutRCtimestamps) respectively. The linear interpolation of the error is shown in (\protect \linedashed), and the maximum and minimum value of the error in (\protect \linedotted), both in their respective color. The maximum absolute error is reduced by a factor 4. Observe that the available error data is non-equidistant in time (inset). Furthermore, the time instances of the available error data are trial-variant. }
%	\caption{Converged error at the timestamps of the intermittent sampling MBFRC design (\protect \markerMBFRCIOHardware), as compared to the initial error (\protect \markerMBFRCTraditionalHardware), i.e., the error without MBFRC, for five basis functions suppressing each roller and the drivetrain. The linear interpolation of the error is shown in (\protect \markerMBFRCdashed) in the respective color.}
	\label{fig:error_time}
\end{figure}

\begin{figure}[!ht]
	\includegraphics[width=\linewidth]{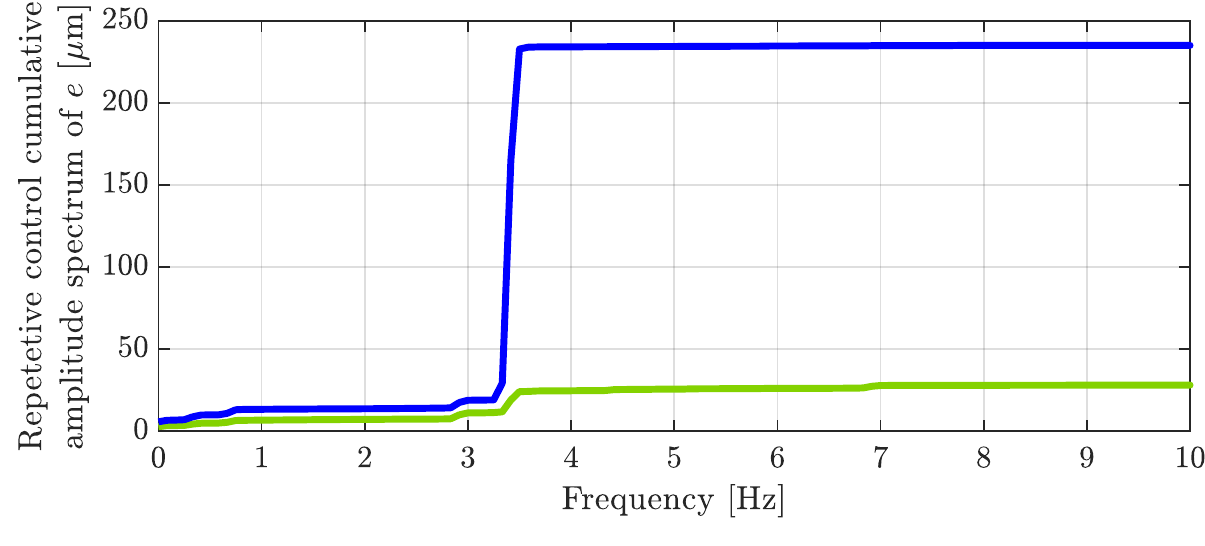}
	\vspace{-2em}	
	\caption{Cumulative amplitude spectrum of the converged error (\protect \linewithRC) and initial error (\protect \linewithoutRC) of the intermittent sampling MBFRC design, constructed from the linearly interpolated equidistant error. The amplitude of the disturbance associated with the drivetrain at $3.5$ Hz is reduced by a factor 10 by the MBFRC design.}
%	\caption{Amplitude spectrum of the converged error of the intermittent sampling MBFRC design (\protect \markerMBFRCIOHardwareLine), as compared to the amplitude spectrum without RC (\protect \markerMBFRCTraditionalHardwareLine). This spectrum is constructed from the interpolated error. The amplitude of the targeted disturbance frequency $f=3.4585$, related to the drivetrain, is reduced by a factor $10$.}
	\label{fig:error_freq}
	\vspace{-1em}
\end{figure}

Fig. \ref{fig:error_time} and \ref{fig:error_freq} show the initial and converged error in time and frequency domain respectively. The MBFRC reduces the maximum absolute error by a factor 4 compared to a situation without RC, based on non-equidistant and trial-variant error data. Furthermore, the amplitude of the error at the drivetrain disturbance frequency of $3.5$ Hz is reduced by a factor 10. 
%The remaining power of the error is mostly associated with frequencies that are not suppressed due to system specific considerations.

\section{Conclusion}
In this paper, a repetitive control framework is introduced that directly enables implementation of RC for a broad range of applications with intermittent sampling characteristics, such as systems that employ optical encoders and networked systems with package loss. Stability conditions are derived for any measurement realization based on small-gain and passivity analysis. The conditions are converted into a straightforward design framework based on loop-shaping and identified FRF data that explicitly allows for addressing uncertainty, closely resembling common RC applications for non-intermittent sampling. The RC framework is experimentally validated through implementation on an industrial printbelt system at Canon Production Printing.

\bibliographystyle{IEEEtran}
\bibliography{IEEEabrv,bibliography.bib}

\end{document}